# Flare: Architecture for rapid and easy development of Internet-based Applications

Shashank Shekhar
School of Computing Sciences, VIT University
shashank.shekhar.alpha@gmail.com

Mohit Soni
School of Computing Sciences, VIT University
mohitsoni.mail@gmail.com

NVSN Kalyan Chakravarthy
School of Computing Sciences, VIT University
nvsnkalyan@gmail.com

## ABSTRACT
We propose an architecture, Flare, that is a structured and easy way to develop applications rapidly, in a multitude of languages, which make use of online storage of data and management of users. The architecture eliminates the need for server-side programming in most cases, creation and management of online database storage servers, re-creation of user management schemes and writing a lot of unnecessary code for accessing different web-based services using their APIs. A Web API provides a common API for various web-based services like Blogger [2], Wordpress, MSN Live, Facebook [3] etc. Access Libraries provided for major programming languages and platforms make it easy to develop applications using the Flare Web Service. We demonstrate a simple micro-blogging service developed using these APIs in two modes: a graphical browser-based mode, and a command-line mode in C++, which provide two different interfaces to the same account and data.

## Categories and Subject Descriptors
K.6.3 [**Management of Computing and Information Systems**]: Software Management – *software development*

H.3.4 [**Information Storage and Retrieval**]: Systems and Software – *distributed systems*

H.3.3 [**Information Storage and Retrieval**]: Information Storage and Retrieval – *clustering, search process*

K.6.5 [**Management of Computing and Information Systems**]: Security and Protection – *authentication*

## 1. MOTIVATION
The Internet has become a ubiquitous entity and the development of new and innovative web-based services has led to an explosion in the amount of content available. Technologies like AJAX, and Rich Internet Application providers like Adobe Flex, and JavaFX have enabled the development of a new breed of innovative and rich applications. These applications harness the storage and computational power of the web, and bring it to the desktop and mobile users. However, there still remains a significant scope for improvement in the techniques and frameworks used for developing such applications. There is a need to simplify the process of writing code for Internet-enabled applications, and also to bring the same ease of use to all major programming languages and platforms. To date, each application uses different sets of classes for accessing web storage services like Amazon S3 or SimpleDB [1] or Box (or manage their own database servers), manage their own users, and write separate access classes for each of the web-based services used like Blogger, Wordpress, Gmail etc. This can get quite cumbersome and results in significantly increased time for development of applications.

Another challenge is that the data generated by applications and users is rapid and evolving in nature. Maintaining suitable storage systems to cater to everybody's computing needs requires careful selection of appropriate storage architectures. As the average Internet access speed across most of the world increases, the Internet is becoming increasingly real-time, often up to a single key-stroke. Coping with such real-time applications requiring very high rates of access, and performance is very important.

The need to simplify the tools and infrastructure for Internet-based applications, and to bridge the gap between the web and traditionally non-browser programming languages is the driving motivation of this architecture. The architecture proposes a simple four-layered model for developing Internet-based applications, with the majority of the repetitive and non-core work being in the bottom three layers, leaving the application developer free to focus on rapidly prototyping and developing new ideas.

## 2. FLARE ARCHITECTURE

### 2.1 Overall Architecture

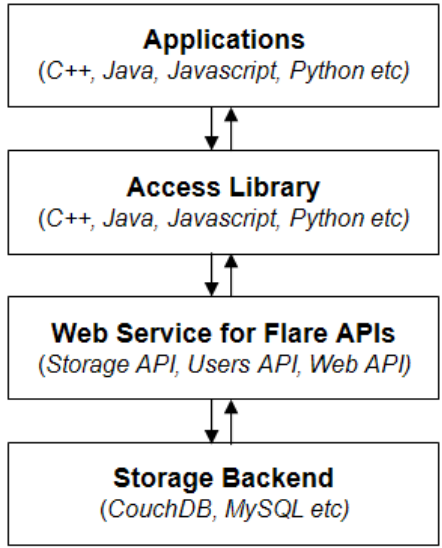

**Figure 1: The Flare architecture**

Flare uses a four-tier architecture that uses open protocols (SOAP [5], REST [6]) for communication. The simplified view of the Flare architecture is given in Figure 1. The top-most layer is formed by Applications that can be written in a multitude of languages and platforms and target hosts (desktop, browser, mobile). The applications layer uses the next layer formed by Access Libraries that provide access to the web service exposed by Flare-enabled servers. The third layer implements the Web service running on Flare-enabled servers for managing the application & user data, and routing data access calls to and from the required external web services. The fourth layer is the Storage backend that provides a high-performance, and indexed storage system for the Flare Web Service.

### 2.2 API Interfaces

The Flare Web Service provides web methods grouped under three APIs: Storage API, Users API, and Web API. These APIs allow apps to store and manage data and users, and access different web services in a unified fashion without requiring any server-side programming on the part of the developer. The Storage API allows the application to easily store and manage data that is accessible across the Internet without requiring database management. The Users API provides the basic operations needed for managing users, and allows applications to manage users specifically meant for that application. The Web API provides a set of common APIs to the developer to easily access web-based services that are similar to each other.

### 2.3 Storage Architecture

Flare can be used with a number of different databases, forming the bottom-most layer in Figure 1. But the conditions that must be met by the storage solution are:

- High performance
- High availability
- Scalability
- Indexed
- Data redundancy
- Error recovery

Apache CouchDB provides a simple and elegant solution and also meets the above conditions. Using CouchDB eliminates the need to model the inherent schema-less data in this service to a schema-oriented database like MySQL. But the architecture permits use of other databases like ThruDB [4], MySQL etc, and both Schema-oriented and Schema-less by changing the wrapper class between Layer 3 (Flare Web Service) and Layer 4 (Storage Backend).

## 3. IMPLEMENATION

The Flare API is exposed as a Web Service supporting SOAP and RESTful techniques of communication. The web methods are grouped into three APIs:

- Storage API
- Users API
- Web API

### 3.1 Storage API

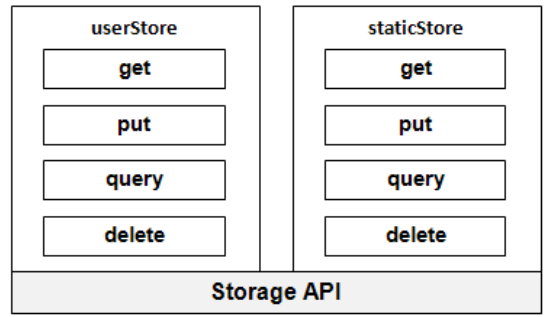

**Figure 2: Storage API**

The Storage API is used for creating globally accessible applications in any of the major languages and permits them to persist and share data between application instances across the Internet. It eliminates the need for server-side programming and database management on the developers' part. It makes it trivially easy to get, put, query/update and delete data from within the code and keeping it specific for each user account of the application. Thus the developer only needs to worry about the Presentation and Application logic.

Two types of storage mechanisms are provided by the Storage API: the **userStore** and **staticStore**. The userStore is a per-user storage provided for applications that store information particular to a user, like his high scores in a gaming application, or his updates in a micro-blogging service built on Flare etc. The data stored is by default *private,* but desired items can be made *public* to enable other users to view the data without authentication. The staticStore is useful for storing data that is accessible across all the instances of the application, regardless of which user is logged in. For e.g. a blogging service built on Flare could use the staticStore to keep track of the ten most recent blog posts made by its users. This data is **static** to all the instances of the application, so a user could see the ten most recent posts made by users of that service across the web.

## 3.2 Users API

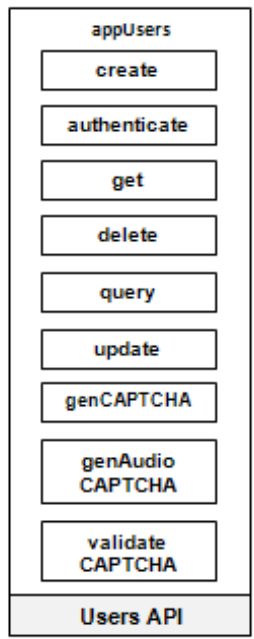

**Figure 3: Users API**

The Users API provides a simple interface for permitting applications to create and manage users. It eliminates the need to redo authentication algorithms, user data storage schemas and server-side programming on the part of the developers as they can store user data and authenticate them from the Flare servers. The API provides flexibility to developers by permitting the storage of as many data key-value pairs about users as required by the application, without imposing a structure on what is stored. This is required because every application differs in what it needs to store about its users. For e.g., social networks built on Flare might need personal and professional information, while simple games might need just the real name of the user, apart from username and password which are the only mandatory fields enforced by the Users API.

## 3.3 Web API

The Web API provides a unified API for web services that are similar to each other. For e.g. all Blogging services can be accessed and used programmatically using the same API exposed by Flare, which implements the code for communicating with the various underlying services like Blogger, Wordpress etc. The Web API categorizes web services into 9 broad groups:

- Blogging
- Email
- Instant Messaging
- Social Networks
- Feeds
- Offbeat devices (SMS, VoIP etc)
- Maps
- Search
- Music, Videos, Photos and other media services

Thus there are nine different APIs providing a minimum common feature specification for each group thereby requiring the developer to work with just one common API for each group. The APIs also provide a mechanism for the developer to access extra features provided by a particular web service that the common API for the group doesn't provide. Thus it becomes easy for developers to build complex mash-ups and provide support for various services in their applications using this API.

## 3.4 Access Libraries

The APIs exposed by Flare will be accessed by libraries ported to each of the major languages and platforms. These libraries will enable desktop applications, mobile applications and web-based Javascript applications to use the storage and unified web APIs in a standard way. Developers could also write multiple interfaces for accessing the same application. For e.g. a Web-based micro-blogging application in Javascript, and a C++ based command-line version of the same application, both using the same userStore item thus providing different interfaces to perform the same actions. All communications with the Web service should use the HTTPS protocol to enforce security [7].

## 4. DEMONSTRATION SCENARIO

This demonstration explains in brief the essential code fragments of *Flitter*, a simple micro-blogging service in two modes: a graphical, browser-based interface in Javascript, and a command-line interface in C++. Both modes provide access to the same account and data.

## 4.1 Javascript version

The Javascript version, shown in Figure 4, uses the Flare library from the server and registers the application along with a developer key with the statements at lines 43 and 44.

```
43  flare.dev.useKey("F92KLFEW5434TR4H");
44  flare.dev.registerApp("flitterApp");
.
.
65  flare.users.authenticate($("#username").val(),
 $("#password").val(), function(r) {
.
.
76  flare.userStore.db.get(["post"], {count: 10}, function
(data) { ...
.
.
87  postData['post']  = {value: $("#post_text").val(),
       access: "public"};
88  flare.userStore.db.put(postData, "posts", function
(data, textStatus) { ...
.
.
102 flare.userStore.db.get(["post"], {userID: otherUID,
count: 10}, function(data) { ...
```

**Figure 4: Code snippets from Fitter.js**

The authentication of the user is performed by line 65 that passes an argument '`r`' to the callback function that is either false, signifying authentication failure, or a valid `userID`. This `userID` is used along with the password in all future transactions. The recent 10 posts written by the user is fetched at line 76. The Javascript library for Flare decodes the XML response sent by the server and sends a key-value map as the argument '`data`' to the callback function.

New posts are stored online by the code at line 87 and 88. This stores the data as *public* in the `userStore` in the item marked by the `userID` and the `appID`.

The other screen takes the username as the input, and gets the ten latest posts by the specified user at line 102. This verifies the visibility mode of the data requested, and returns the ten posts that can be shown on the screen.

## 4.2 C++ version

The C++ version, shown in Figure 5, uses a library enclosed in `flare.dll`, and includes the necessary header file on line 2.

```
2  #include "flare.h"
.
.
43 flare::userStore::db::get("['posts']", "{'count':10}",
   show_my_posts);
.
.
77 flare::userStore::db::get("['posts']", "{'userID':'" +
   otherUID + "','count':10}", show_other_posts);
```

**Figure 5: Code snippets from Flitter.cpp**

The remaining syntax for performing the tasks is almost identical, except with the replacement of anonymous functions (that were used in the Javascript version) with function pointers and the scope resolution operator (`::`) being used for namespace access as in line 43. This snippet calls a developer-defined function `show_my_posts()` and passes to it the received data. The application can also show posts by other users with the code at line 77, using syntax similar to the Javascript version. The C++ library includes a JSON parser (that permits single quote for ease of use), and parameters are passed as JSON strings as shown in line 43 and 77.

## 5. APPLICABILITY

While our demonstration application was a simple one, the architecture and API can be successfully implemented on Web servers to permit an elegant way for developers to simplify development. A few, specialized systems that require custom storage systems and data management schemas might not be suitable candidates for using the Flare architecture. But this addresses the needs of a very large base of web and application developers looking for simpler ways to create Internet-enabled applications that work on multiple devices and platforms.

The kind of applications that can be developed using the Flare APIs is really diverse owing to the lack of enforcement of structure, and a generic design. Applications can be simple and complex, including simple games and mash-ups to social networks, blogging services and service aggregators. However, there remain significant challenges in keeping the system secure and highly available, and more development and design is needed to refine the architecture further to cover deeper issues.